\documentclass[sigconf,natbib=true]{acmart}

\usepackage{amsmath,amssymb,amsfonts}
\usepackage{algorithmic}
\usepackage{graphicx}
\usepackage{textcomp}
\usepackage{booktabs}
\usepackage{xcolor}
\usepackage{threeparttable}
\usepackage{float}
\usepackage{multirow}
\usepackage{bm}
\usepackage{CJK}
\usepackage{indentfirst}
\usepackage{cases}
\usepackage{ulem}
\usepackage{caption}
\usepackage{subfigure}
\usepackage{makecell}
\usepackage[ruled]{algorithm2e}
\usepackage{hyperref}
\usepackage{hyperref}
\hypersetup{
	colorlinks=true,
	linkcolor=blue,
	citecolor=blue,
	filecolor=magenta,
	urlcolor=cyan,
}
\usepackage{xcolor}

\usepackage{ragged2e}

\AtBeginDocument{%
  \providecommand\BibTeX{{%
    \normalfont B\kern-0.5em{\scshape i\kern-0.25em b}\kern-0.8em\TeX}}}

\copyrightyear{2023}
\acmYear{2023}
\setcopyright{acmlicensed}\acmConference[SIGIR '23]{Proceedings of the 46th International ACM SIGIR Conference on Research and Development in Information Retrieval}{July 23--27, 2023}{Taipei, Taiwan}
\acmBooktitle{Proceedings of the 46th International ACM SIGIR Conference on Research and Development in Information Retrieval (SIGIR '23), July 23--27, 2023, Taipei, Taiwan}
\acmPrice{15.00}
\acmDOI{10.1145/3539618.3591752}
\acmISBN{978-1-4503-9408-6/23/07}

\begin{document}
\begin{sloppypar}

\title{Prompt Learning for News Recommendation}

\author{Zizhuo Zhang}
\affiliation{%
	\institution{School of Electronic Information and Communications, Huazhong University of Science and Technology}
	\streetaddress{Huazhong University of Science and Technology}
	\city{Wuhan}
	\country{China}}
\email{zhangzizhuo@hust.edu.cn}

\author{Bang Wang}
\affiliation{%
	\institution{School of Electronic Information and Communications, Huazhong University of Science and Technology}
	\streetaddress{Huazhong University of Science and Technology}
	\city{Wuhan}
	\country{China}}
\email{wangbang@hust.edu.cn}

\begin{abstract}
	Some recent \textit{news recommendation} (NR) methods introduce a Pre-trained Language Model (PLM) to encode news representation by following the vanilla pre-train and fine-tune paradigm with carefully-designed recommendation-specific neural networks and objective functions. Due to the inconsistent task objective with that of PLM, we argue that their modeling paradigm has not well exploited the abundant semantic information and linguistic knowledge embedded in the pre-training process. Recently, the pre-train, prompt, and predict paradigm, called \textit{prompt learning}, has achieved many successes in natural language processing domain. In this paper, we make the first trial of this new paradigm to develop a \textit{Prompt Learning for News Recommendation} (Prompt4NR) framework, which transforms the task of predicting whether a user would click a candidate news as a cloze-style mask-prediction task. Specifically, we design a series of prompt templates, including discrete, continuous, and hybrid templates, and construct their corresponding answer spaces to examine the proposed Prompt4NR framework. Furthermore, we use the prompt ensembling to integrate predictions from multiple prompt templates. Extensive experiments on the MIND dataset validate the effectiveness of our Prompt4NR with a set of new benchmark results.
\end{abstract}





\maketitle

%\footnote{https://news.google.com/}

\section{Introduction}
Nowadays, online news platforms such as Google News has become a vital portal for people to efficient acquire daily information~\cite{das:2007:WWW}. News recommendation (NR), as a filtering tool to alleviate the information overload problem~\cite{mo:2018:FGCS,liu:2017:SIGIR:EBSN}, can effectively help users to find their mostly interested news articles among the huge amount of news~\cite{wu:2022:TOIS:Survey,okura:2017:SIGKDD}.

\par
\begin{figure*}[t]
	\centering
	\includegraphics[width=0.99\textwidth]{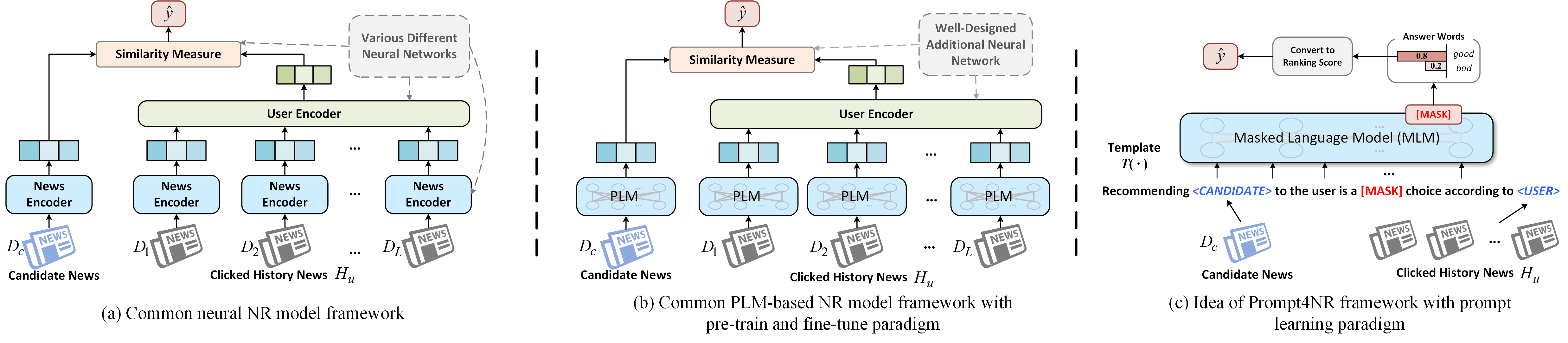}
	\caption{The differences of neural NR models, pre-train and fine-tune PLM-based models, and our Prompt4NR framework.}
	\label{Fig:Neural-PLM-Prompt}
\end{figure*}

Most existing neural NR methods have mainly focused on designing various ingenious neural networks to encode news' and users' representations~\cite{wang:2018:WWW:DKN,an:2019:ACL:LSTUR,wu:2019:EMNLP:NRMS,wu:2019:SIGKDD:NPA,zhu:2019:AAAI:DAN,zhang:2019:ICDM,wang:2020:ACL:FIM,zhang:2021:SIGIR:AMM,qi:2021:ACL:HieRec,qi:2022:SIGIR:FUM,kim:2022:CIKM,li:2022:ACL:MINER}. We summarize the common approaches in these models by Figure~\ref{Fig:Neural-PLM-Prompt}(a), where the core modules are news encoder, user encoder and similarity measure. In the literature, these modules have been implemented via different neural networks. For example, Wu et al.~\cite{wu:2019:EMNLP:NRMS} leverage a multi-head self-attention network as news encoder and user encoder. Wang et al.~\cite{wang:2020:ACL:FIM} propose to learn a kind of hierarchical multi-level news representation via stacked dilated convolutions for fine-grained matching. In these neural models, the static word embeddings (e.g., Word2Vec~\cite{mikolov:2013:Arxiv:Word2vec} and Glove~\cite{pennington:2014:EMNLP:Glove}) are mostly adopted as initializations in model training, which mainly focuses on mining in-domain information in a NR dataset, yet ignoring the abundant semantic and linguistic information from real-world large-scale corpus.

%*******************************
%*******************************

\par
Some recent methods take one step further to introduce a \textit{pre-trained language model} (PLM) for learning news representations~\cite{wu:2021:SIGIR:PLM4NR,zhang:2021:IJCAI:UNBERT,jia:2021:SIGIR:RMBERT,yu:2021:Arxiv:TinyNewsRec,xiao:2022:SIGKDD,bi:2022:ACL:MTRec}. We summarize their common approaches by Figure~\ref{Fig:Neural-PLM-Prompt}(b), where the vanilla \textit{pre-train and fine-tune} paradigm~\cite{liu:2021:PromptSurvey} is used to adapt the downstream NR task. In this paradigm, a PLM is only used as news encoder, yet another neural network is designed for encoding users. A NR-specific objective function is used to train the whole model. Although these methods show promising performance improvements, they have not well exploited the abundant encyclopedia-like knowledge in large-scale PLMs due to the inconsistency between the downstream NR objective and the PLM training objective.

\par
Recently, a novel \textit{pre-train, prompt and predict} paradigm named \textit{prompt learning} has exhibited remarkable successes in many applications in the \textit{natural language processing} (NLP) domain~\cite{dai:2022:EMNLP:Prompt4EAE,xiang:2022:CoLing:Prompt4IDDR}. The basic of this new paradigm is to reformulate a downstream task into the PLM training task via designing task-related \textit{prompt template} and \textit{answer words space}~\cite{liu:2021:PromptSurvey,geng:2022:Recsys:P5}. For its promising potentials, we are also interested to examine the applicability and effectiveness of the prompt learning paradigm in the NR domain. To the best of our knowledge, this paper makes the first trial and proposes a \textit{Prompt Learning for News Recommendation} (Prompt4NR) framework.

\par
Figure~\ref{Fig:Neural-PLM-Prompt}(c) illustrates the design philosophy of our Prompt4NR framework, where we propose to convert a NR task into a cloze-style mask-prediction task. In particular, given the click history of a user $H_u=\{D_1,D_2,...,D_L\}$ and a candidate news $D_c$, we first convert $H_u$ into a sentence, denoted as \texttt{<USER>} so as to encode the user interests from history. We also convert $D_c$ into a sentence, denoted as \texttt{<CANDIDATE>} for a candidate news. Then we design a \textit{prompt template} $T(\text{\texttt{<USER>}}, \text{\texttt{<CANDIDATE>}})$ to concatenate the two sentences into another sentence with a \textsf{[MASK]} token. After passing a Masked Language Model (MLM), an answer space corresponding to the template is designed to predict the \textsf{[MASK]}, which is then converted to a ranking score to determine whether this candidate news should be recommended to the user.

\par
In this paper, we present a series of investigations on how the core design issues of Prompt4NR would impact on the recommendation performance. They include the design issues of (1) Template: What kind of templates is more suitable for integrating news data and user behaviors in the NR domain? (2) Verbalizer: How to map the recommendation labels to answer words for \textsf{[MASK]} prediction? (3) Ensemble: Can we integrate the advantages of different templates to boost the performance? For prompt template, we design three types of templates, including discrete, continuous, and hybrid templates, from the considerations of four mostly interested aspects between \texttt{<USER>} and \texttt{<CANDIDATE>} for the NR task, including the \textit{semantic relevance}, \textit{user emotion}, \textit{user action} and \textit{recommendation utility}. For verbalizer, we construct a binary answer space with two answer words with opposite senses according to the template, which corresponds to the real label of whether the user clicks the candidate news. For ensemble, we use multi-prompt ensembling to make a decision fusion of predictions from different templates. We design extensive experiments on the wide-used MIND dataset~\cite{wu:2020:ACL:MIND}. A set of new benchmark results validates the effectiveness of our Prompt4NR framework in terms of better recommendation performance over the state-of-the-art competitors.

\section{Related Work}
In this section, we review the existing related work on news recommendation and prompt learning.
\subsection{News Recommendation}
Various neural networks have been widely developed to encode news and user representations for the NR task~\cite{park:2017:CIKM,wang:2018:WWW:DKN,khattar:2018:CIKM,wu:2019:IJCAI:NAML,wu:2019:EMNLP:NRHUB,an:2019:ACL:LSTUR,wu:2019:EMNLP:NRMS,wu:2019:ACL:TANR,wu:2019:SIGKDD:NPA,zhu:2019:AAAI:DAN,zhang:2019:ICDM,wang:2020:ACL:FIM,liu:2020:IJCAI:Hypernews,wu:2020:IJCAI:CPRS,zhang:2021:SIGIR:AMM,qi:2022:SIGIR:Fastformer,zhang:2021:WWW,qi:2021:ACL:HieRec,wang:2021:CIKM:PENR,qi:2021:ACL:PPRec,wu:2022:ACL:TBOS,wu:2022:WWW:FeedRec,qi:2022:SIGIR:FUM,kim:2022:CIKM,li:2022:ACL:MINER,wu:2022:SIGIR:MMRec,qi:2022:SIGIR:CAUM,Gong:2022:SIGIR}. For example, Wu et al.~\cite{wu:2019:EMNLP:NRMS} propose the NRMS to adopt the multi-head attention network as news and user encoders. Wang et al.~\cite{wang:2020:ACL:FIM} propose the FIM to make the fine-grained matching via stacked dilated convolutions. Qi et al.~\cite{qi:2021:ACL:HieRec} propose the HieRec that represents each user as a hierarchical interest tree based on topics and subtopics of news. Li et al.~\cite{li:2022:ACL:MINER} propose the MINER to capture a user's multiple interests representations. Though effective, these shallow neural networks lack the ability to understand the deep linguistic and semantic information in news texts. Also their models learn knowledge only from supervised signals in the NR task, ignoring the potential benefits from real-world large-scale corpora.

\par
Recently, with great success of PLM in the NLP domain, some methods have incorporated the PLM for the NR task and have achieved substantial improvements~\cite{wu:2021:SIGIR:PLM4NR,zhang:2021:IJCAI:UNBERT,jia:2021:SIGIR:RMBERT,yu:2021:Arxiv:TinyNewsRec,xiao:2022:SIGKDD,bi:2022:ACL:MTRec}. For example, Wu et al.~\cite{wu:2021:SIGIR:PLM4NR} propose an intuitionistic PLM-based NR framework, which directly instantiates the news encoder in existing neural NR models using a PLM. Bi et al.~\cite{bi:2022:ACL:MTRec} use the multi-task learning framework to enhance the capability of BERT to encode multi-field information of news such as category and entity. These above methods all follow the vanilla pre-train and fine-tune paradigm with a NR-specific objective. Despite achieving some performance improvements, such paradigm may be suboptimal for exploiting knowledge learned from the pre-training process for the NR task, due to inconsistent objective with that of PLM~\cite{liu:2021:PromptSurvey}.

\subsection{Prompt Learning}
Prompt learning is a novel learning strategy upon a PLM, which converts a downstream task into the form of \textsf{[MASK]} prediction using the PLM by adding template to the input texts. How to design prompt template is the core component of prompt learning. Generally, the types of templates can be categorized as discrete, continuous and hybrid templates~\cite{liu:2021:PromptSurvey}. Discrete templates consist of existing natural words, relying heavily on human experiences. For example, Petroni et al.~\cite{petroni:2019:EMNLP} manually design template to probe knowledge in LMs. Schick et al.~\cite{Schick:2021:EACL} propose a semi-supervised training procedure PET to reformulate specific tasks as cloze-style tasks. Although discrete template has succeeded in many tasks, handcrafted templates may be with costs and not globally optimal. To overcome such issues, continuous and hybrid templates introduce learnable prompt tokens to automatically search the templates, such as the AutoPrompt~\cite{shin:2020:EMNLP:autoprompt}, Prefix Tuning~\cite{li:2021:ACL:prefix}, P-Tuning~\cite{liu:2021:P-Tuning}, P-Tuningv2~\cite{liu:2021:P-Tuningv2} and etc. In this paper, we focus on how to exploit the prompt learning for the NR task. We propose and experiment a set of discrete, continuous and hybrid templates to investigate different design considerations.

\begin{figure}[t]
	\centering
	\includegraphics[width=\linewidth]{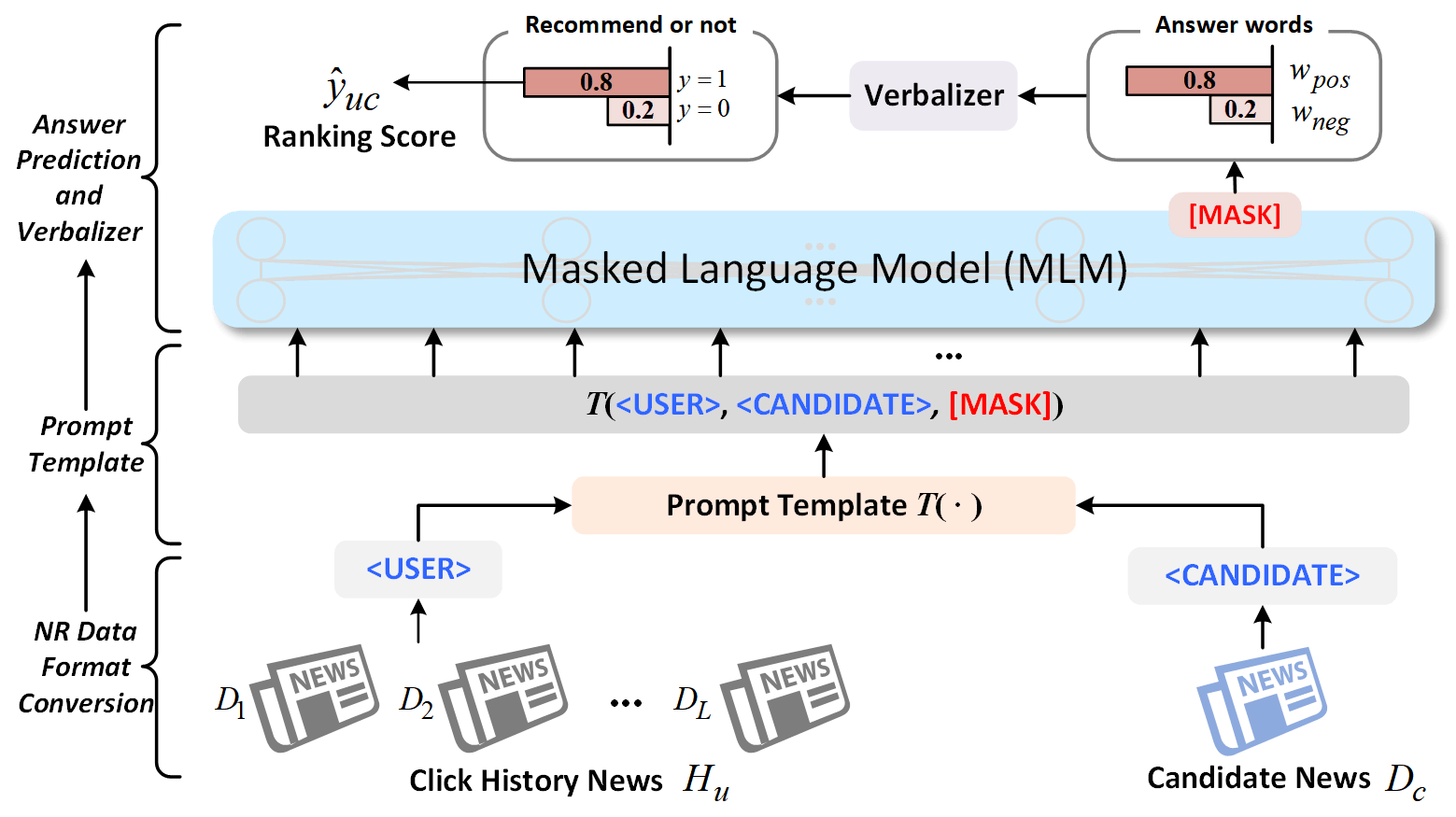}
	\caption{Illustration of the Prompt4NR Framework.}
	\label{Fig:OverallModel}
\end{figure}

\section{The Prompt4NR Framework}
We first present the problem statement of NR and next introduce our Prompt4NR framework. We then explain the training strategy and multi-prompt ensembling.

\subsection{The NR Task and Prompt Overview}
\label{Sec:NR-ProblemStatement}
Denote the $\mathcal{U}$ and $\mathcal{D}$ as the user set and news set, respectively. Each news $D\in\mathcal{D}$ is mainly equipped with its title $Title=\{w_1, w_2, ..., w_M\}$ that is a sequence of words, where $M$ is the number of words. Given a user $u$'s click history $H_u=\{D_1, D_2, ..., D_L\}$ containing $L$ clicked news and a candidate news $D_c$, the news recommendation (NR) task aims at predicting a ranking score $\hat{y}_{uc}$ to estimate the probability that the user $u$ would click the candidate news $D_c$. The candidate news with the highest score will be recommended to the user.

\par
Fig.~\ref{Fig:OverallModel} illustrates our Prompt4NR framework, which contains three main modules: (1) NR data format conversion (Section~\ref{Sec:<USER>and<CANDIDATE>}); (2) Prompt template (Section~\ref{Sec:PromptTemplate}); (3) Answer prediction and verbalizer (Section~\ref{Sec:AnswerAndVerbalizer}). We explain their details as follows:

\subsection{NR Data Format Conversion}
\label{Sec:<USER>and<CANDIDATE>}
Given a click history $H_u$ and candidate news $D_c$, we convert them into a natural language sentence to adapt to the subsequent prompt learning paradigm, denoted as \texttt{<USER>} and \texttt{<CANDIDATE>}, respectively. For the \texttt{<USER>}, we concatenate the news titles of a user history $H_u$, where a \textit{virtual token} \textsf{[NCLS]} is added at the beginning of each title to segment each clicked news. For the \texttt{<CANDIDATE>}, we adopt the title of the candidate news $D_c$. We formally denote them by:
\begin{flushleft}
	\qquad \qquad \texttt{<USER>} $\leftarrow$ \textsf{[NCLS]} $Title_1$ ... \textsf{[NCLS]} $Title_L$ \\
	\qquad \qquad \texttt{<CANDIDATE>} $\leftarrow$ $Title_c$
\end{flushleft}
where $\{Title_1, ..., Title_L\}$ correspond to news titles in $H_u=\{D_1, D_2, ..., D_L\}$. Intuitively, the \texttt{<USER>} can be regarded as a summary of the user $u$'s area of interest, and the \texttt{<CANDIDATE>} condenses the core textual semantics of a candidate news. Both of them serve as the input textual data for subsequent prompt templates.

\begin{table*}[t]
	\caption{Prompt templates designed in this paper, including discrete, continuous and hybrid templates.}
	\tabcolsep=0.1cm
	\renewcommand\arraystretch{1.5}
	\resizebox{2\columnwidth}{!}{
		\begin{tabular}{|c|c|c|c|}
			\hline
			\textbf{Types} & \textbf{Perspectives} & \textbf{Templates $\bm{T}(\text{\texttt{\textcolor{blue}{<USER>}}}, \text{\texttt{\textcolor{blue}{<CANDIDATE>}}}, \text{\textsf{\textcolor{red}{[MASK]}}})$} & \textbf{Answer Words} \\
			\hline
			\multirow{4}*{\makecell[c]{\textbf{Discrete} \\ \textbf{Template}}} & Relevance & \texttt{\textcolor{blue}{<CANDIDATE>}} is \textcolor{red}{\textsf{[MASK]}} to \texttt{\textcolor{blue}{<USER>}} & related/unrelated \\
			\cline{2-4}
			~ & Emotion & The user feels \textcolor{red}{\textsf{[MASK]}} about \texttt{\textcolor{blue}{<CANDIDATE>}} according to his area of interest \texttt{\textcolor{blue}{<USER>}} & interesting/boring \\
			\cline{2-4}
			~ & Action & User: \texttt{\textcolor{blue}{<USER>}} \textsf{[SEP]} News: \texttt{\textcolor{blue}{<CANDIDATE>}} \textsf{[SEP]} Does the user click the news? \textcolor{red}{\textsf{[MASK]}} & yes/no \\
			\cline{2-4}
			~ & Utility & Recommending \texttt{\textcolor{blue}{<CANDIDATE>}} to the user is a \textcolor{red}{\textsf{[MASK]}} choice according to \texttt{\textcolor{blue}{<USER>}} & good/bad \\
			\hline
			\multirow{4}*{\makecell[c]{\textbf{Continuous} \\ \textbf{Template}}} & Relevance & $[Q_1]...[Q_{n_2}]$ \texttt{\textcolor{blue}{<CANDIDATE>}} $[M_1]...[M_{n_3}]$ \textsf{\textcolor{red}{[MASK]}} $[P_1]...[P_{n_1}]$ \texttt{\textcolor{blue}{<USER>}} & related/unrelated \\
			\cline{2-4}
			~ & Emotion & $[M_1]...[M_{n_3}]$ \textsf{\textcolor{red}{[MASK]}} $[Q_1]...[Q_{n_2}]$ \texttt{\textcolor{blue}{<CANDIDATE>}}  $[P_1]...[P_{n_1}]$ \texttt{\textcolor{blue}{<USER>}} & interesting/boring \\
			\cline{2-4}
			~ & Action & $[P_1]...[P_{n_1}]$ \texttt{\textcolor{blue}{<USER>}} \textsf{[SEP]} $[Q_1]...[Q_{n_2}]$ \texttt{\textcolor{blue}{<CANDIDATE>}} \textsf{[SEP]} $[M_1]...[M_{n_3}]$ \textcolor{red}{\textsf{[MASK]}} & yes/no \\
			\cline{2-4}
			~ & Utility & $[Q_1]...[Q_{n_2}]$ \texttt{\textcolor{blue}{<CANDIDATE>}} $[M_1]...[M_{n_3}]$ \textcolor{red}{\textsf{[MASK]}} $[P_1]...[P_{n_1}]$ \texttt{\textcolor{blue}{<USER>}} & good/bad \\
			\hline
			\multirow{4}*{\makecell[c]{\textbf{Hybrid} \\ \textbf{Template}}} & Relevance & $[P_1]...[P_{n_1}]$ \texttt{\textcolor{blue}{<USER>}} \textsf{[SEP]} $[Q_1]...[Q_{n_2}]$ \texttt{\textcolor{blue}{<CANDIDATE>}} \textsf{[SEP]} This news is \textsf{\textcolor{red}{[MASK]}} to the user's area of interest & related/unrelated \\
			\cline{2-4}
			~ & Emotion & $[P_1]...[P_{n_1}]$ \texttt{\textcolor{blue}{<USER>}} \textsf{[SEP]} $[Q_1]...[Q_{n_2}]$ \texttt{\textcolor{blue}{<CANDIDATE>}} \textsf{[SEP]} The user feels \textsf{\textcolor{red}{[MASK]}} about the news & interesting/boring \\
			\cline{2-4}
			~ & Action & $[P_1]...[P_{n_1}]$ \texttt{\textcolor{blue}{<USER>}} \textsf{[SEP]} $[Q_1]...[Q_{n_2}]$ \texttt{\textcolor{blue}{<CANDIDATE>}} \textsf{[SEP]} Does the user click the news? \textcolor{red}{\textsf{[MASK]}} & yes/no \\
			\cline{2-4}
			~ & Utility & $[P_1]...[P_{n_1}]$ \texttt{\textcolor{blue}{<USER>}} \textsf{[SEP]} $[Q_1]...[Q_{n_2}]$ \texttt{\textcolor{blue}{<CANDIDATE>}} \textsf{[SEP]} Recommending the news to the user is a \textcolor{red}{\textsf{[MASK]}} choice & good/bad \\
			\hline
	\end{tabular}}
	\label{Tble:PromptTemplates}
\end{table*}

\subsection{Prompt Template}
\label{Sec:PromptTemplate}
As the core component of Prompt4NR, prompt template $T(\cdot)$ wraps the input data $(\text{\texttt{<USER>}}, \text{\texttt{<CANDIDATE>}})$ to convert the NR task as a cloze-style task to predict the \textsf{[MASK]}:
\begin{equation}
	x_{prompt} = T(\text{\texttt{<USER>}}, \text{\texttt{<CANDIDATE>}}, \text{\textsf{[MASK]}})
\end{equation}
As the first work exploiting prompt learning for the NR task, we are interested to know what kind of prompt templates would most benefit recommendation performance. To this end, we design three types of templates, including discrete, continuous and hybrid templates, from different perspectives of how to capture the matching signals between a user and a candidate news. Table~\ref{Tble:PromptTemplates} summarizes our designed prompt templates. We next introduce the design ideas behind these templates.

\subsubsection{\textbf{Discrete Templates}}
As the most common type of template engineering in prompt learning, discrete templates form the input data via human-interpretable natural language, which requires some prior experiential knowledge. We design four discrete templates from four different considerations, where each corresponds to one way to measure the matching signals between a user's interest and a candidate news. That is, we explore which style of \textsf{[MASK]} cloze as the similarity measurement is suitable for the NR task.

\par
$\bullet$ \textbf{Semantic Relevance:} This is to examine whether the related news contents are the core motivation for a user to read a news. In other words, whether a user is with a kind of persistent interests to some particular topics and contents. To this end, we convert the NR task to determine the relevance between \texttt{<CANDIDATE>} and \texttt{<USER>}, and the answer words are chosen as \textit{"related"} and \textit{"unrelated"}.

\par
$\bullet$ \textbf{User Emotion:} This is to investigate whether users' emotional reactions to news would be the most impacting factor. In other words, a user chooses to read a news, as if the news could mostly satisfy a user emotional needs. We use the emotion words \textit{"interesting"} and \textit{"boring"} as the answers to estimate the user emotional reaction to the \texttt{<CANDIDATE>}.

\par
$\bullet$ \textbf{User action:} This is to study whether a MLM can directly serve as a click predictor. In other words, interest guides action and action reflects interest. After telling the MLM the \texttt{<USER>} and \texttt{<CANDIDATE>}, we let the MLM directly predict whether the user would click on the news, and the answer words are \textit{"yes"} and \textit{"no"}.

\par
$\bullet$ \textbf{Recommendation utility:} This is to explore whether a MLM can itself make a judgement about the potential merits and demerits of recommending a candidate news, that is, the utility of making such a recommendation. To this end, we prompts the MLM with an utilization question, and the answer words are \textit{"good"} and \textit{"bad"} as the recommendation utility predictions.

\par
The above templates, though seemingly with only a few differences on the template sentences and answer words, are expected to exploit the semantics and linguistics knowledge embedded in a large encyclopedia-like PLM through predefined natural sentences and preselected answer words. On the one hand, we note that this is the core philosophy of the prompt learning paradigm, that is, predicting the probability of an answer word from the PLM vocabulary, as if the task-specific input sentences have been inserted into the large corpus for training the PLM. On the other hand, such manually designed templates, though with well-designed natural sentences, are obviously not exhaustive for all possible cases. As such, we may use some virtual tokens to search a few more cases in a template, that is, a continuous template.

\subsubsection{\textbf{Continuous Template}}
Table~\ref{Tble:PromptTemplates} presents our design of four continuous templates, each corresponding to one discrete template. We add some virtual learnable tokens in front of the \texttt{<USER>}, \texttt{<CANDIDATE>} and \textsf{[MASK]}, respectively, denoted as $[P_1]...[P_{n_1}]$, $[Q_1]...[Q_{n_2}]$ and $[M_1]...[M_{n_3}]$, where $n_1, n_2, n_3$ are numbers of virtual tokens. As for the answer words and token position setting, we refer to previous discrete templates. Although continuous templates provide the model more freedom, the embeddings of these virtual tokens are randomly initialized, which may introduce some ambiguities, leading to under-utilization of the PLM knowledge. We further design a kind of hybrid templates, trying to combine the advantages of both discrete and continuous templates.

\subsubsection{\textbf{Hybrid Template}}
In a hybrid template, we preserve those virtual tokens $[P_i]$ and $[Q_j]$ in front of \texttt{<USER>} and \texttt{<CANDIDATE>}, with the aim of automatically searching for the appropriate formats to present these information to PLM. We replace those virtual tokens $[M_k]$ with a natural language with the \textsf{[MASK]} token that is used for answer prediction. As presented in Table~\ref{Tble:PromptTemplates}, we still design four representative natural sentences each corresponding to one of our design considerations. A hybrid template is hence composed of a continuous template, a \textsf[SEP] token and a natural sentence. Compared with those continuous templates, we argue that such hybrid templates can enjoy the virtual tokens for more choices by the continuous templates, yet still bearing natural sentences for guiding answer directions by the discrete templates.

%In hybrid template, we preserve these virtual tokens $[P_1]...[P_{n_1}]$ and $[Q_1]...[Q_{n_2}]$ in front of \textit{<USER>} and \textit{<CANDIDATE>}. For $[M_1]...[M_{n_3}]$, we use natural language to replace them and concatenate the \textsf{[MASK]} token. As shown in Table~\ref{Tble:PromptTemplates}, we still build the natural language prompts from the previous four perspectives to obtain four hybrid templates. All of them adopt the similar position placement, from left to right, we present the \textit{<USER>} and \textit{<CANDIDATE>}, finally we give out a sentence with a \textsf{[MASK]} token for prediction. Compared with continuous templates, we argue that the these natural language prompts is helpful in inducing the PLM to search the answer words.

\subsection{Answer Prediction And Verbalizer}
\label{Sec:AnswerAndVerbalizer}
Given a click history $H_u$ and a candidate news $D_c$, they correspond to a real label $y\in \{0,1\}$ reflecting whether the user clicks the candidate news ($y=1$) or not ($y=0$). We design a \textit{verbalizer} $v(\cdot)$ to map the labels to two answer words in the PLM vocabulary $\mathcal{W}$ as follows:
\begin{equation}
	v(y) = \left\{
	\begin{aligned}
		w_{pos},\quad&y=1, \\
		w_{neg},\quad&y=0,
	\end{aligned}
	\right.
\end{equation}
where $\mathcal{W}_a=\{w_{pos}, w_{neg}\}\subset \mathcal{W}$ is the answer word space, which can be different according to the used prompt templates. The NR task is converted into a cloze-style task that the pre-trained MLM $\mathcal{M}$ (e.g. BERT~\cite{devlin:2018:Arxiv:Bert}) predicts the probability of answer words to be the \textsf{[MASK]}:
\begin{equation}
	P(y|H_u, D_c) = P_{\mathcal{M}}\left([MASK]=v(y)|x_{prompt}\right),
\end{equation}
where $P_{\mathcal{M}}\left([MASK]=v(y=1)|x_{prompt}\right)$ can be regarded as the confidence of whether to recommend the current candidate news. We use it as the ranking score to form a recommendation list. We note that this paper considers a simple answer space construction with two PLM vocabulary words. We will investigate more complicated answer space construction with more vocabulary words and even virtual answers in our future work.

\subsection{Training}
Compared with the pre-train and fine-tune paradigm to train additional task-specific neural models, our Prompt4NR model only has parameters of PLM to tune. We adopt the cross entropy loss function to train our model:
\begin{equation}
	\mathcal{L} = -\frac{1}{N} \sum_{i=1}^{N} \left[y_i\log P_i + (1-y_i)\log (1-P_i)\right],
\end{equation}
where $y_i$ and $P_i$ are the gold label and predicted probability of the $i$-th training instance, respectively. We use the AdamW~\cite{Loshchilov:2019:ICLR:AdamW} optimizer with L2 regularization for model training.

\subsection{Multi-Prompt Ensembling}
Different templates may have different advantages, as they each focus on a particular design consideration, resulting in their different utilizations of the linguistic and semantic knowledge in a PLM. Multi-prompt ensembling fuses the predictions of individual prompts to boost final decision~\cite{lester:2021:EMNLP,liu:2021:PromptSurvey}. As we have no prior knowledge about which template is better, we simply sum the probability of the positive answer word in each prompt as the final ranking score:
\begin{equation}
	\hat{y} = \sum_{e\in \mathcal{E}} P_e,
\end{equation}
where $P_e$ is the template $e$'s output probability for $w_{pos}$, $\mathcal{E}$ is the set of fused templates. In this paper, we consider two kinds of multi-prompt ensembling. One is to fuse predictions from the same type of templates, where $\mathcal{E}=\{\text{\textit{Relevance, Emotion, Action, Utility}}\}$. That is, a discrete ensembling is to fuse the four discrete templates' predictions. The other is to fuse predictions from different types of templates, named \textit{cross-type ensembling}. %\wbmark{In particular, our cross-type ensembling is to fuse the best performing template in each type. BETTER DELETE}
%% 因为最好的模板不知道，建议删掉这句话

%Different templates may have different advantages and focus on different linguistic and semantic knowledge in PLM. Prompt ensembling uses multiple prompts for an input at inference time to boost the predictions~\cite{lester:2021:EMNLP,liu:2021:PromptSurvey}. We have designed the prompt templates from four perspectives, each of them can make its own prediction, and we integrate their predictions to co-determine the final recommendation results. Specifically, we sum the probability for positive answer word of each prompt as the final ranking score:
%\begin{equation}
%	\hat{y} = P_{Rel} + P_{Emo} + P_{Beh} + P_{Uti}
%\end{equation}
%where $P_*$ is the output probability for $w_{pos}$. We make prompt ensembling under discrete, continuous and hybrid templates respectively. Furthermore, we make a cross-type ensembling by selecting the the best performing templates from discrete, continuous and hybrid prompts respectively.

%\begin{table}[t]
%	\centering
%	\caption{Dataset statistics.}
%	\begin{tabular}{ccccc}
%		\toprule[1pt]
%		\multirow{4}*{MIND} & \# Users & \# News & \# Clicks & \# Impressions \\
%		~ & 94,057 & 65,238 & 347,727 & 230,117 \\
%		\cline{2-5}
%		~ & \multicolumn{2}{c}{Avg. history news} & \multicolumn{2}{c}{Avg. title length} \\
%		~ & \multicolumn{2}{c}{32.46} & \multicolumn{2}{c}{11.52} \\
% 		\bottomrule[1pt]
%	\end{tabular}
%	\label{Tble:Dataset}
%\end{table}

\begin{table}[t]
	\centering
	\caption{Dataset statistics.}
	\begin{tabular}{ccccc}
		\toprule[1pt]
		\multirow{2}*{MIND} & \# Users & \# News & \# Clicks & \# Impressions \\
		~ & 94,057 & 65,238 & 347,727 & 230,117 \\
		\bottomrule[1pt]
	\end{tabular}
	\label{Tble:Dataset}
\end{table}

\section{Experiment Settings}
In this section, we introduce our experimental settings, including dataset, parameter settings, evaluation metrics and baselines.
\subsection{Dataset}
We conduct experiments on the public real-world NR benchmark dataset MIND\footnote{https://msnews.github.io/}~\cite{wu:2020:ACL:MIND}, where users' behaviors are recorded by impressions. An impression log records the clicked and non-clicked news that are displayed to a user at a specific time and his historical news click behaviors before this impression. MIND collects the user behavior logs from October 12 to November 22, 2019 from the Microsoft News platform. Following previous work~\cite{qi:2021:ACL:HieRec,bi:2022:ACL:MTRec}, user data in the first four weeks is used to construct users' history, user data in penultimate week is used as training set and user data in the last week as testing set, we extract 5\% impressions from training set to form the validating set. Table~\ref{Tble:Dataset} summarizes the dataset statistics.

\begin{table}[t]
	\centering
	\caption{The overall comparison of performance results.}
	\begin{threeparttable}
		\renewcommand\arraystretch{1.1}
		\resizebox{1\columnwidth}{!}{
			\begin{tabular}{|c|c|cccc|}
				\hline
				\multicolumn{2}{|c|}{Dataset} & \multicolumn{4}{c|}{MIND} \\
				\hline
				\multicolumn{2}{|c|}{Baselines} & AUC & MRR & NDCG@5 & NDCG@10 \\
				\hline
				\multirow{4}*{\makecell{Neural \\ Methods}} & NPA & 64.65 & 30.01 & 33.14 & 39.47 \\
				~ & LSTUR & 65.87 & 30.78 & 33.95 & 40.15 \\
				~ & NRMS & 65.63 & 30.96 & 34.13 & 40.52 \\
				~ & FIM & 65.34 & 30.64 & 33.61 & 40.16 \\
				\hline
				\multirow{3}*{\makecell{PLM-based \\ Methods}} & BERT-NPA & 67.56 & 31.94 & 35.34 & 41.73 \\
				~ & BERT-LSTUR & 68.28 & 32.58 & 35.99 & 42.32 \\
				~ & BERT-NRMS & 68.60 & 32.97 & 36.55 & 42.78 \\
				\hline
				\hline
				\multicolumn{2}{|c|}{Prompt4NR (BERT)} & AUC & MRR & NDCG@5 & NDCG@10 \\
				\hline
				\multirow{5}*{\makecell{Discrete \\ Template}} & Relevance & 68.77 & 33.42 & 37.20 & 43.36 \\
				~ & Emotion & 68.77 & 33.29 & 37.12 & 43.19 \\
				~ & Action & 68.76 & 33.22 & 37.02 & 43.26 \\
				~ & Utility & \underline{68.94} & \underline{33.62} & \underline{37.47} & \underline{43.61} \\
				\cline{2-6}
				~ & Ensembling & \textbf{69.34} & \textbf{33.76} & \textbf{37.71} & \textbf{43.80} \\
				\hline
				\multirow{5}*{\makecell{Continuous \\ Template}} & Relevance & \underline{69.25} & 33.72 & 37.75 & 43.79 \\
				~ & Emotion & 68.76 & 33.51 & 37.39 & 43.47 \\
				~ & Action & 68.58 & 33.37 & 37.17 & 43.30 \\
				~ & Utility & 69.10 & \underline{33.96} & \underline{37.91} & \underline{43.92} \\
				\cline{2-6}
				~ & Ensembling & \textbf{69.43} & \textbf{34.06} & \textbf{38.11} & \textbf{44.14} \\
				\hline
				\multirow{5}*{\makecell{Hybrid \\ Template}} & Relevance & 68.47 & 33.26 & 37.20 & 43.24 \\
				~ & Emotion & 68.59 & 33.26 & 37.19 & 43.29 \\
				~ & Action & \textbf{69.37} & \textbf{34.02} & \textbf{37.96} & \textbf{44.00} \\
				~ & Utility & 68.79 & 33.45 & 37.35 & 43.49 \\
				\cline{2-6}
				~ & Ensembling & \underline{69.22} & \underline{33.78} & \underline{37.77} & \underline{43.87} \\
				\hline
				\multicolumn{2}{|c|}{Cross-Type Ensembling} & \textbf{\color{blue}{69.64}} & \textbf{\color{blue}{34.26}} & \textbf{\color{blue}{38.30}} & \textbf{\color{blue}{44.33}} \\
				\hline
		\end{tabular}}
		\begin{tablenotes}
			\footnotesize
			\item[1] Boldface with blue indicates the best results in the whole table. In each type of  \\ template, only boldface indicates the best results, the second best is underlined.
			\item[2] The cross-type ensembling fuses the best performings in each type for decision, \\ i.e., $\mathcal{E}=\{\text{\textit{Discrete-Utility, Continuous-Utility, Hybrid-Action}}\}$.
			%,  i.e., $\mathcal{E}=\{\text{\textit{Discrete-Utility, Continuous-Utility, Hybrid-Action}}\}$.
			\item[3] The improvements are significant ($p<0.01$) as validated by student's t-test.
		\end{tablenotes}
	\end{threeparttable}
	\label{Tble:OverallPerformance}
\end{table}

\subsection{Parameter Settings}
In our experiments, we employ the base BERT (12 layers, bert-base-uncased)~\cite{devlin:2018:Arxiv:Bert} implemented by HuggingFace\footnote{https://huggingface.co/} transformers~\cite{wolf:2020:EMNLP:huggingface} as the PLM to experiment our Prompt4NR. We adopt the AdamW optimizer with learning rate lr=2e-5 to train the model. We run the model with 8 NVIDIA RTX-A5000 GPUs in distributed data parallel and the batch size on each of them is set as 16, which is equal to batch size 128 on a single GPU. We apply negative sampling with ratio 4 the same as other baselines. Following previous work~\cite{wu:2021:SIGIR:PLM4NR,wang:2020:ACL:FIM,wu:2019:EMNLP:NRMS}, we adopt a user's most recent 50 clicked news as the user's click history. For each news in history, we set the maximum length of the title to be 10 words; for each candidate news, we set the maximum length of the title to be 20 words. Besides, we use the average AUC, MRR, NDCG@5 and NDCG@10 over all impressions as the evaluation metrics, which are widely used in the NR studies~\cite{wu:2021:SIGIR:PLM4NR,wang:2020:ACL:FIM,qi:2021:ACL:HieRec}. All
hyperparameters are adjusted on the validating set. We have released the source code at \href{https://github.com/resistzzz/Prompt4NR}{https://github.com/resistzzz/Prompt4NR}.

\subsection{Baselines}
We compare our methods with following competitors, which can be categorized into neural NR models and BERT-enhanced models:
\par
\textbf{Neural NR models:} Various neural networks have been specially designed for the NR task, including: (1) \textit{NPA}~\cite{wu:2019:SIGKDD:NPA} uses the personalized attention network on words and history clicked news for news and user representation learning. (2) \textit{LSTUR}~\cite{an:2019:ACL:LSTUR} captures a user's short-term and long-term interests via a GRU network and the user's embedding respectively. (3) \textit{NRMS}~\cite{wu:2019:EMNLP:NRMS} uses the multi-head self-attention to learn news and user representations. (4) \textit{FIM}~\cite{wang:2020:ACL:FIM} adopts a hierarchical dilated convolution network to encode news representation from multiple grained views.

\par
\textbf{BERT-enhanced models:} Wu et al.~\cite{wu:2021:SIGIR:PLM4NR} adopt the BERT as the news encoder on several of the above methods. We reproduce three baselines here for comparison, denoted as \textit{BERT-NPA}, \textit{BERT-LSTUR} and \textit{BERT-NRMS}, for which we only replace their news encoder with BERT and keep the other neural parts (e.g., user encoder) unchanged.

\section{Experiment Results}

\begin{figure}[t]
	\centering
	\subfigure[Impact of $n_1, n_2, n_3$ in \textit{Continuous-Utility} template.]{
		\includegraphics[width=0.85\linewidth]{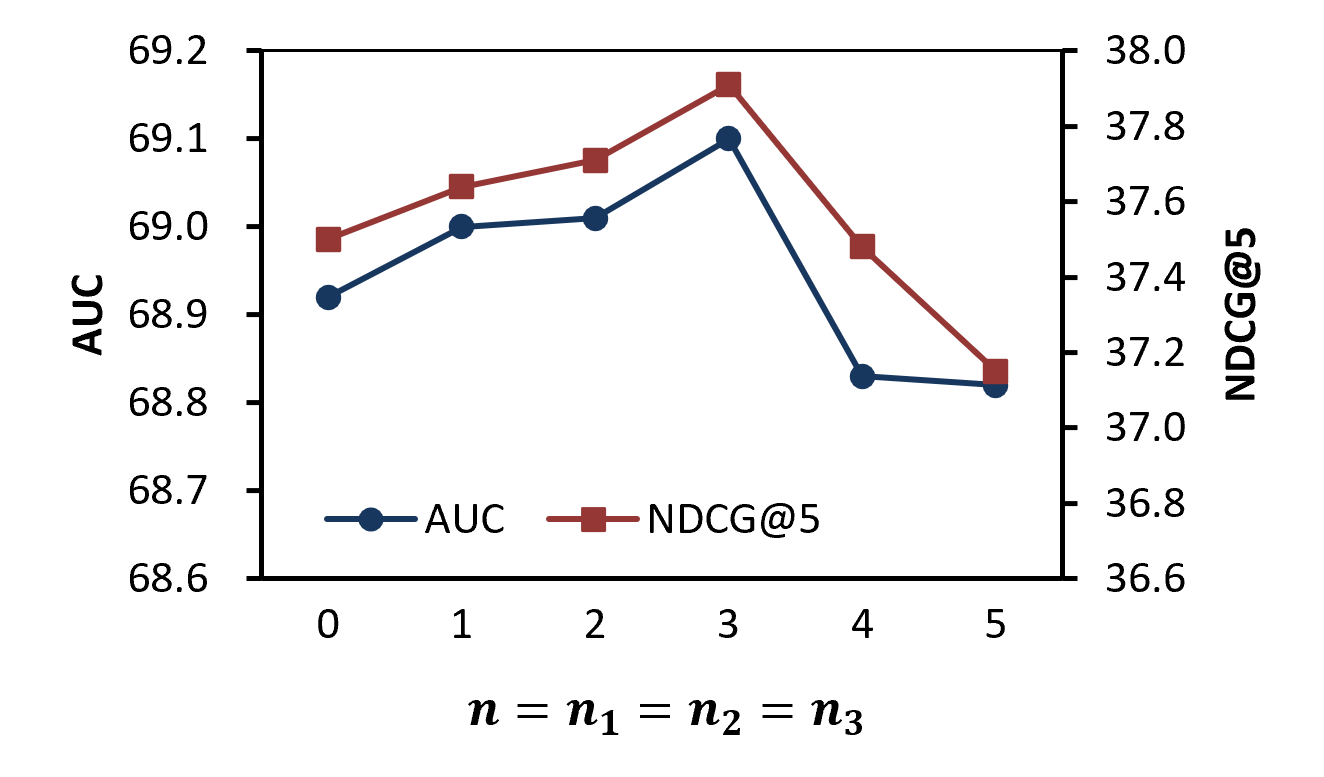}
		\label{Fig:Conti-n}}
	%	\hspace{0.1cm}
	\subfigure[Impact of $n_1, n_2$ in \textit{Hybrid-Action} template.]{
		\includegraphics[width=0.85\linewidth]{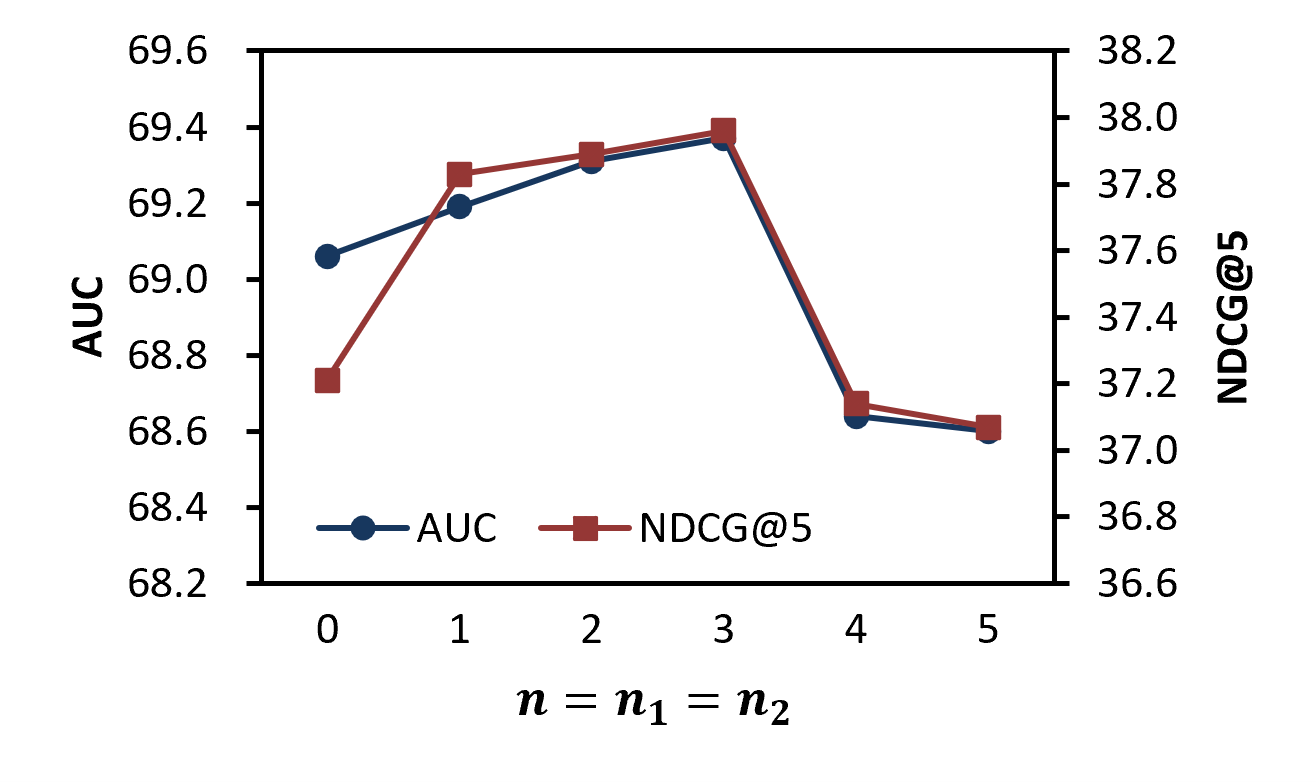}
		\label{Fig:Hybrid-n}}
	\caption{Impact of the number of virtual tokens in the continuous and hybrid templates.}
	\label{Fig:Hyperparameters}
\end{figure}

\subsection{Main Experiment Results}
Table~\ref{Tble:OverallPerformance} summarizes the comparison of our Prompt4NR with these state-of-the-art baselines. The boldface with blue indicates the best results in the whole table. In each type of template, only boldface indicates the best result in its kind, while the second best is underlined. From these results, we have following observations:

\par
Firstly, the models using BERT as news encoder are consistently better than those neural models using shallow neural networks as news encoder, e.g., BERT-NRMS outperforms NRMS. This indicates that the pre-trained BERT brings benefits to news encoding, which is attributed to the rich semantic and linguistic knowledge learned by BERT during the pre-training process.

\par
Secondly, our Prompt4NR methods based on prompt-learning paradigm, either discrete, continuous or hybrid templates, almost outperform those BERT-enhanced models based on the pre-train, fine-tune paradigm. Especially, the metrics of MRR, NDCG@5 and NDCG@10 all surpass 33.00, 37.00 and 43.00, respectively, and AUC reults on some templates surpass 69.00. This reflects the superiority of the prompt learning paradigm for developing knowledge embedded in pre-trained BERT to support the NR task.

\par
Thirdly, three different types of templates achieve comparable performances, and the order of their bests is "\textit{Discrete-Utility} $<$ \textit{Continuous-Utility} $<$ \textit{Hybrid-Action}", in which \textit{Hybrid-Action} is the best in all of the one-single kind templates. Furthermore, the performance gaps within the four discrete templates are relatively smaller than those within the continuous and hybrid templates. This is because of those learnable virtual tokens in continuous and hybrid templates that bring more potentials yet with more variations.

\begin{figure}[t]
	\centering
	\includegraphics[width=0.9\columnwidth]{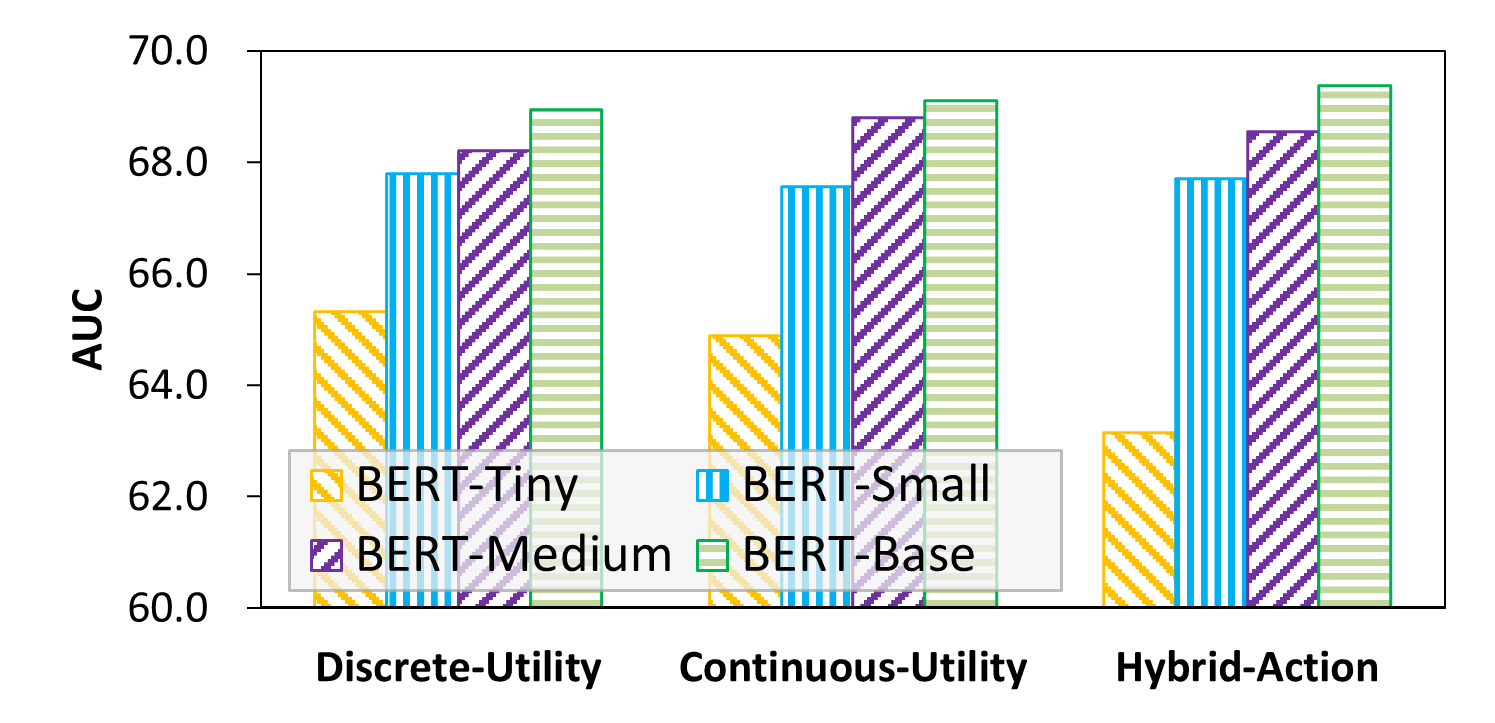}
	\includegraphics[width=0.9\columnwidth]{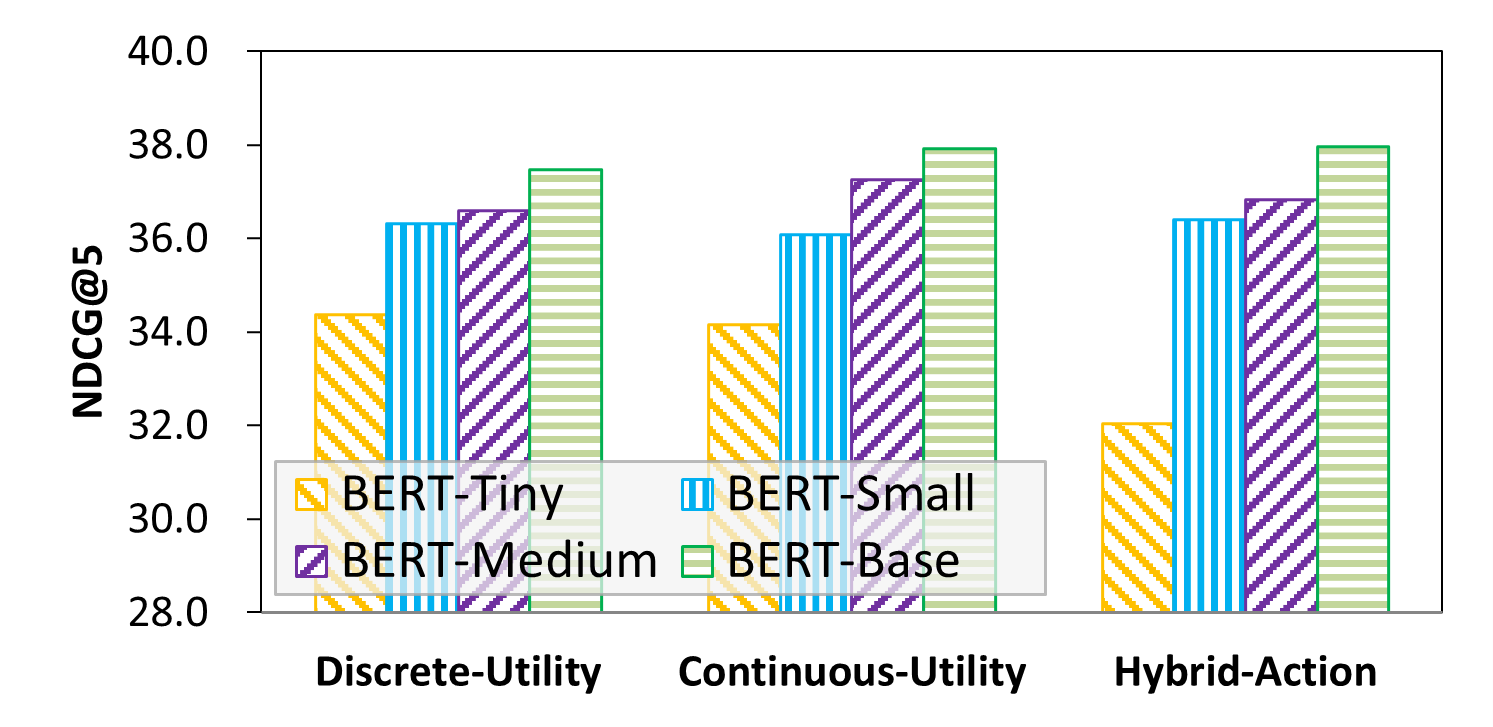}
	\caption{Impact of the BERT scale.}
	\label{Fig:BERTScale}
\end{figure}

\begin{table}[t]
	\centering
	\caption{Performance of using different PLMs.}
	\begin{threeparttable}
		\renewcommand\arraystretch{1.1}
		\begin{tabular}{c|cccc}
			\hline
			Discrete-Utility & AUC & MRR & NDCG@5 & NDCG@10 \\
			\hline
			BERT & 68.94 & 33.62 & 37.47 & 43.61 \\
			RoBerta & \textbf{69.20} & \textbf{34.00} & \textbf{37.77} & \textbf{43.96} \\
			DeBerta & \underline{69.01} & \underline{33.98} & \underline{37.67} & \underline{43.74} \\
			\hline
		\end{tabular}
	\end{threeparttable}
	\label{Tble:PerformanceOfDifferentPLMs}
\end{table}

\par
Finally, performing multi-prompt ensembling obviously improves the recommendation performance than using only a one-single prompt on the discrete and continuous templates. This indicates the effectiveness of prompt ensembling to fuse multi-prompts for better decision. On the hybrid templates, the ensembling result is better than the relevance, emotion and utility template, but worse than the action template. This suggests that the prerequisite for prompt ensemble to work is that the performance gap between the individual fused prompts should not be too large. In addition, cross-type ensembling, which fuses the three best in each type, i.e., \textit{Discrete-Utility}, \textit{Continuous-Utility} and \textit{Hybrid-Action}, achieves the best performance. This reflects that prompt ensembling should not be limited by template type, and that cross-type ensembling may be a good choice in some cases.

\par
By the way, we notice that the three bests in each template type are from the \textit{Utility}, \textit{Utility} and \textit{Action} prompt, respectively. In the four considerations of prompt templates, the \textit{Action} and \textit{Utility} prompts are directly related to the recommendation task; While the \textit{Relevance} and \textit{Emotion} prompts are not directly related. The \textit{Relevance} and \textit{Emotion} prompts have transformed the interest matching by inferring the semantic relevance between news texts and predicting users' emotional reactions, respectively. We suspect those templates directly associated with the recommendation properties other than using another kind of underlying inference may be more suitable for the NR task.

%The \textit{Relevance} prompt infers users' reading interests first and then makes matching next, and the \textit{Emotion} predicts users' emotional reactions first. From the experiments results, we argue that those templates directly associated with the recommendation task other than using another kind of underlying inference would be more suitable for the NR task.
%This also suggests to further investigate the potentials of prompt learning in our future work.

%We hope that our a series of templates and experiments can inspire subsequent related research work.

\subsection{Hyperparameter Analysis}
The core hyperparameters of our Prompt4NR are the numbers of virtual tokens in continuous templates (i.e. $n_1, n_2, n_3$) and hybrid templates (i.e. $n_1, n_2$). We notice that there are so many combinations of them taking different values, even if their values vary only in a small range. We adopt a coarse hyperparameter tuning strategy, that is, we let $n_1, n_2, n_3$ take the same value for tuning, denoted as $n=n_1=n_2=n_3$.

\par
In particular, we vary $n$ in $\{0,1,2,3,4,5\}$, where we notice that when $n=0$ without any tokens is called NullPrompt~\cite{logan:2022:ACL:NullPrompt}. Considering the paper length, we only present the tuning results of the best two, i.e., \textit{Continuous-Utility} and \textit{Hybrid-Action}, the trends of the others are similar. Figure~\ref{Fig:Hyperparameters} plots the recommendation performance against different values of $n$. We find that the trends are the same for both continuous and hybrid templates. That is, as $n$ increases, the performance first improves and then decreases. This indicates that only a few of virtual tokens are enough for prompting: Too fewer may lack guidance to organize information to a PLM; While too many may suffer from more ambiguities because randomly initialized virtual tokens lack pre-training knowledge. Furthermore, we notice that even Null Prompt (i.e. $n=0$), the AUC can reach 68.9, which has surpassed those BERT-enhanced methods. This again reflects the superiority of the prompt learning paradigm on the NR task, with its ability to better exploit the PLM to capture the potential relationships between news texts.

\subsection{Impact of PLM}
\subsubsection{\textbf{Impact of The PLM's Scale}}
We investigate the impact of using different scales of BERT, i.e., BERT-Tiny (2 layers), BERT-Small (4 layers), BERT-Medium (8 layers) and BERT-Base (12 layers). We perform different scales of BERT on the best template in each type. Figure~\ref{Fig:BERTScale} plots the results of AUC and NDCG@5. We observe that using larger PLMs with more parameters and deeper layers usually leads to better recommendation performance. This is not unexpected, because a larger PLM usually has stronger abilities to capture the deep linguistic and semantic information of news texts, and the performance of prompt learning also depends on the PLM's such capability. We reasonably guess the performance can be further improved by using larger BERT (e.g., BERT-Large with 24 layers). However, we believe that we should strike a balance between the recommendation performance and the model scale, as an oversized PLM may not be suitable for real application scenarios.

\subsubsection{\textbf{Impact of Different PLMs}}
We replace the BERT with other two popular PLMs, viz., the RoBerta~\cite{liu:2019:Roberta} and DeBerta~\cite{he:2021:ICLR:deberta}, to investigate the impact of using different PLMs. Both the RoBerta and DeBerta are the improved version of the BERT\footnote{BERT, RoBerta and DeBerta all adopt the base version with 12 layers. Their model name on huggingface are \{bert-base-uncased, roberta-base, deberta-base\}.}. The RoBerta removes the next sentence prediction and is pre-trained on a larger corpus with larger batch size; The DeBerta introduces a disentangled attention mechanism and an enhanced mask decoder to improve BERT. Table~\ref{Tble:PerformanceOfDifferentPLMs} presents the experiment results. We observe that both the RoBerta and DeBerta outperform the BERT, and RoBerta achieves the best results. This suggests that the downstream NR task can benefit from the improvements made in the pre-training process.

\begin{figure}[t]
	\centering
	\includegraphics[width=0.85\columnwidth]{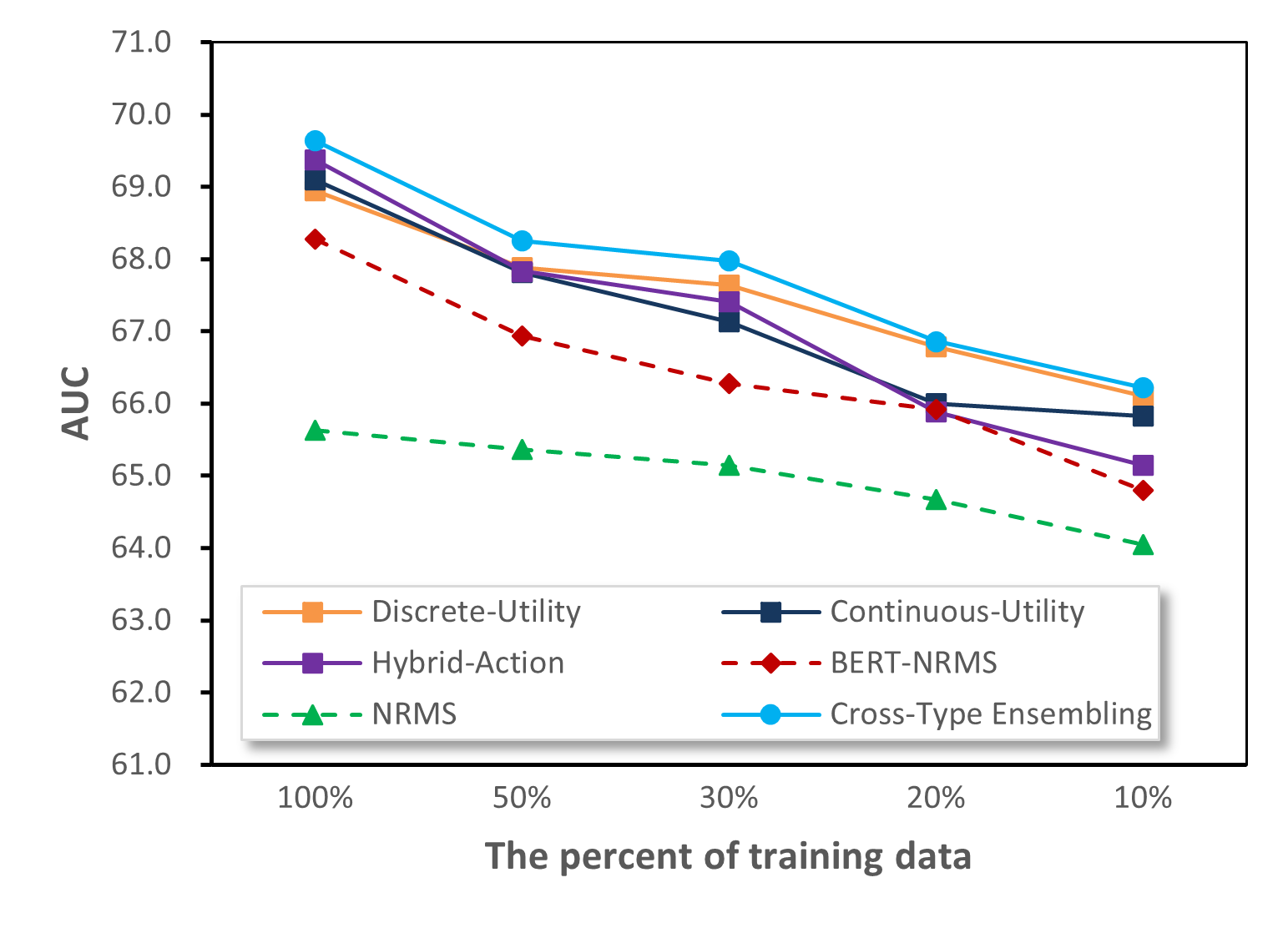}
	\includegraphics[width=0.85\columnwidth]{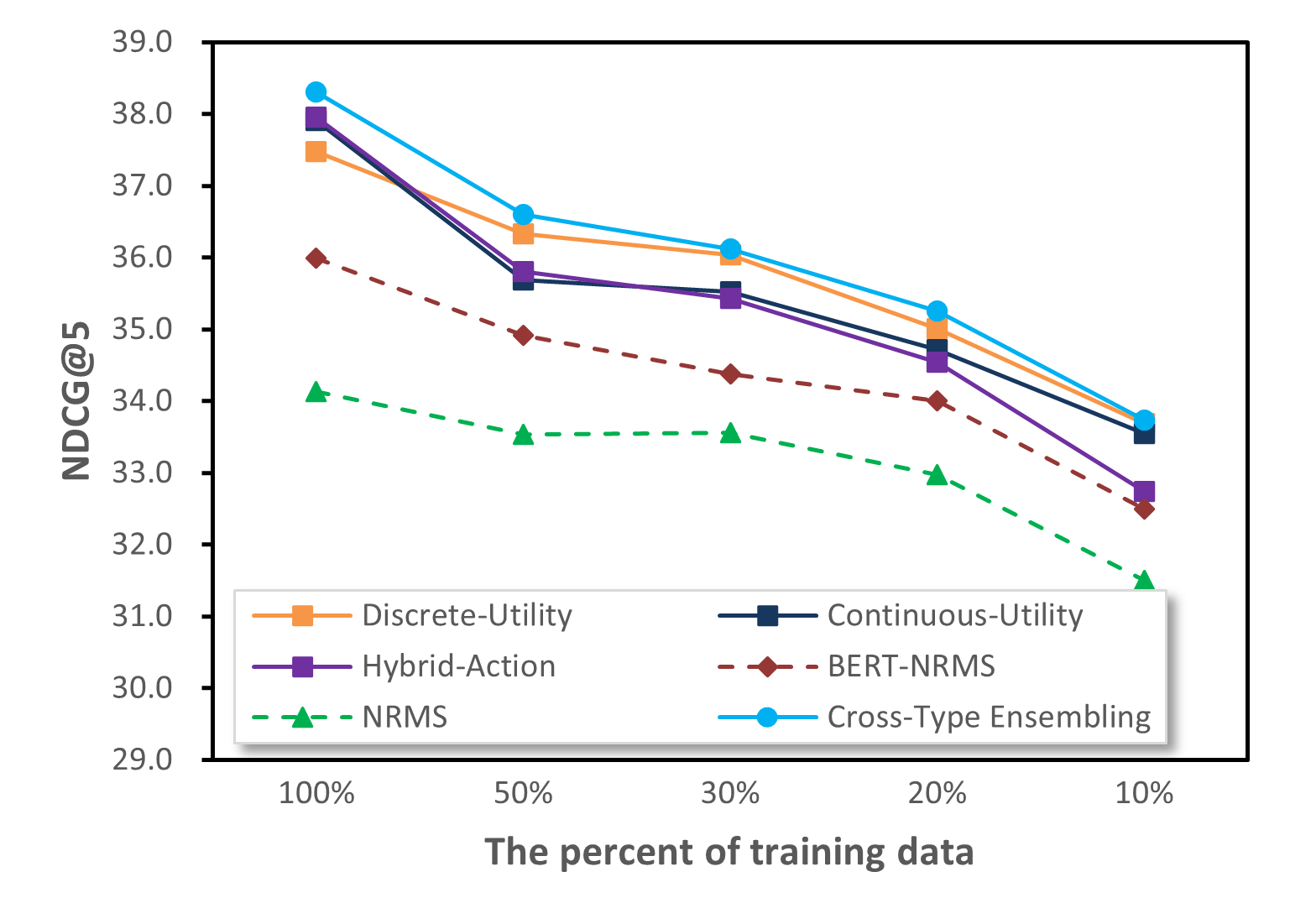}
	\caption{Performance comparison of few-shot learning.}
	\label{Fig:FewShot}
\end{figure}

\subsection{Performance of Few-shot Learning}
Some researchers have reported that the prompt learning paradigm has some robustness under few-shot scenarios for some NLP tasks, such as text classification~\cite{wang:2021:EMNLP:transprompt}, implicit discourse relation recognition~\cite{xiang:2022:CoLing:Prompt4IDDR}. We would also like to examine our Prompt4NR and competitors under few-shot learning scenarios. We adopt the down-sampling to gradually reduce the training set, while keeping the validating set and testing set unchanged.

\begin{figure*}[t]
	\centering
	\includegraphics[width=2\columnwidth, height=0.33\textwidth]{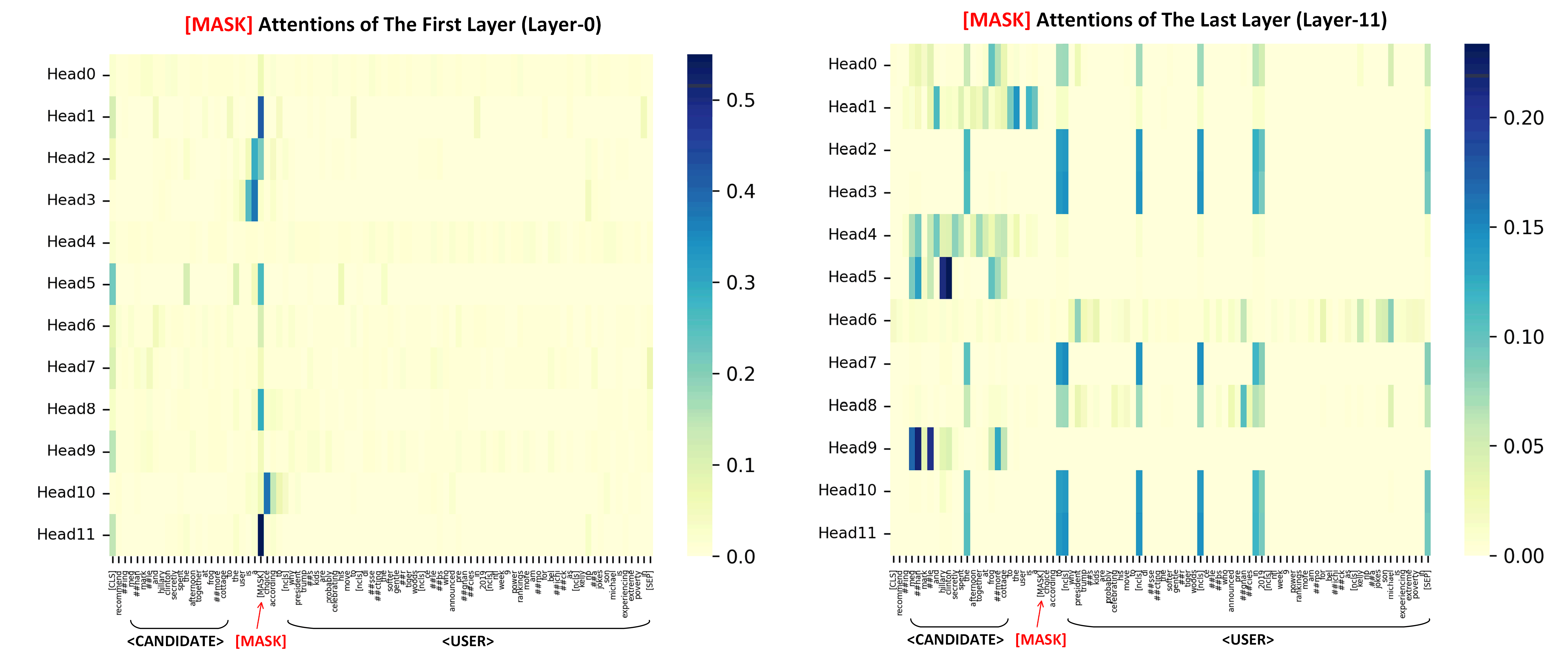}
	\caption{Visualization of the BERT's 12-heads attention weights of the \textsf{[MASK]} token for a sampled instance: The left is the attention weights of the first layer; The right is the last layer. The \textit{X}-axis is the input sentence, the \textit{Y}-axis is the 12 heads of BERT, and the BERT has been trained by using the \textit{Discrete-Utility} template.}
	\label{Fig:Visualization}
\end{figure*}

\par
Figure~\ref{Fig:FewShot} summarizes the performance comparison of few-shot learning, where dashed lines denote two competitors NRMS and BERT-NRMS, and full lines denote three one-single templates and a cross-type ensembling of our Prompt4NR. Not surprisingly, both our Prompt4NR and competitors suffer from performance degradation as training data decreases. We observe that the line of NRMS is always at the bottom; The line of our three one-single templates are higher than that of the BERT-NRMS in most of cases; The line of cross-ensembling is higher than all the others. The results show that the PLM has stronger capability to handle few-shot scenarios than shallow neural networks, viz., the prompt learning is more robust than the vanilla pre-train and fine-tune paradigm. The prompt ensembling strategy can further boost the robustness. Furthermore, it is observed that when decreasing training set, the \textit{Discrete-Utility} is in an upper hand position more often relative to \textit{Continuous-Utility} and \textit{Hybrid-Action}. This may be due to that those virtual tokens in the continuous and hybrid templates are not sufficiently trained as the training data decreases.

\subsection{Visualization}
To examine what BERT has learned in our Prompt4NR framework, we visualize the attention weights in BERT of a random case in Figure~\ref{Fig:Visualization}. Since BERT has 12-layers and each layer has 12-heads, we visualize the \textsf{[MASK]}'s attention weights of the first layer and the last layer to help us understand the learning process. In the first layer, the attentions mainly focus on these tokens in the area around \textsf{[MASK]}. In the last layer, we observe that the focus is different for different heads. Some heads still focus on the area around \textsf{[MASK]}, e.g., Head1. Some heads pay more attention on the area of \texttt{<CANDIDATE>}, e.g., Head4, Head5 and Head9. Some heads pay more attention on the area of \texttt{<USER>}, where \{Head6, Head8\} and \{Head2, Head3, Head7, Head10, Head11\} focus on different points; Head6 and Head8 have wide attentions on specific tokens in \texttt{<USER>}, we called token-level attention; While Head2, Head3, Head7, Head10 and Head11 focus especially on the areas around of several \texttt{[NCLS]} tokens, we called news-level attention, because the virtual token \texttt{[NCLS]} at the beginning of each historical news may have ability to represent its following news. The attentions of Head0 is dispersed throughout the sentence. In summary, compared to the first layer, the last layer has clearer attention points, and each head makes its contributions. This reflects the fact that the BERT, by being trained in our Prompt4NR framework, can discriminately use available information for the \textsf{[MASK]} prediction and recommendation.

\begin{figure*}[t]
	\centering
	\includegraphics[width=2\columnwidth, height=0.62\textheight]{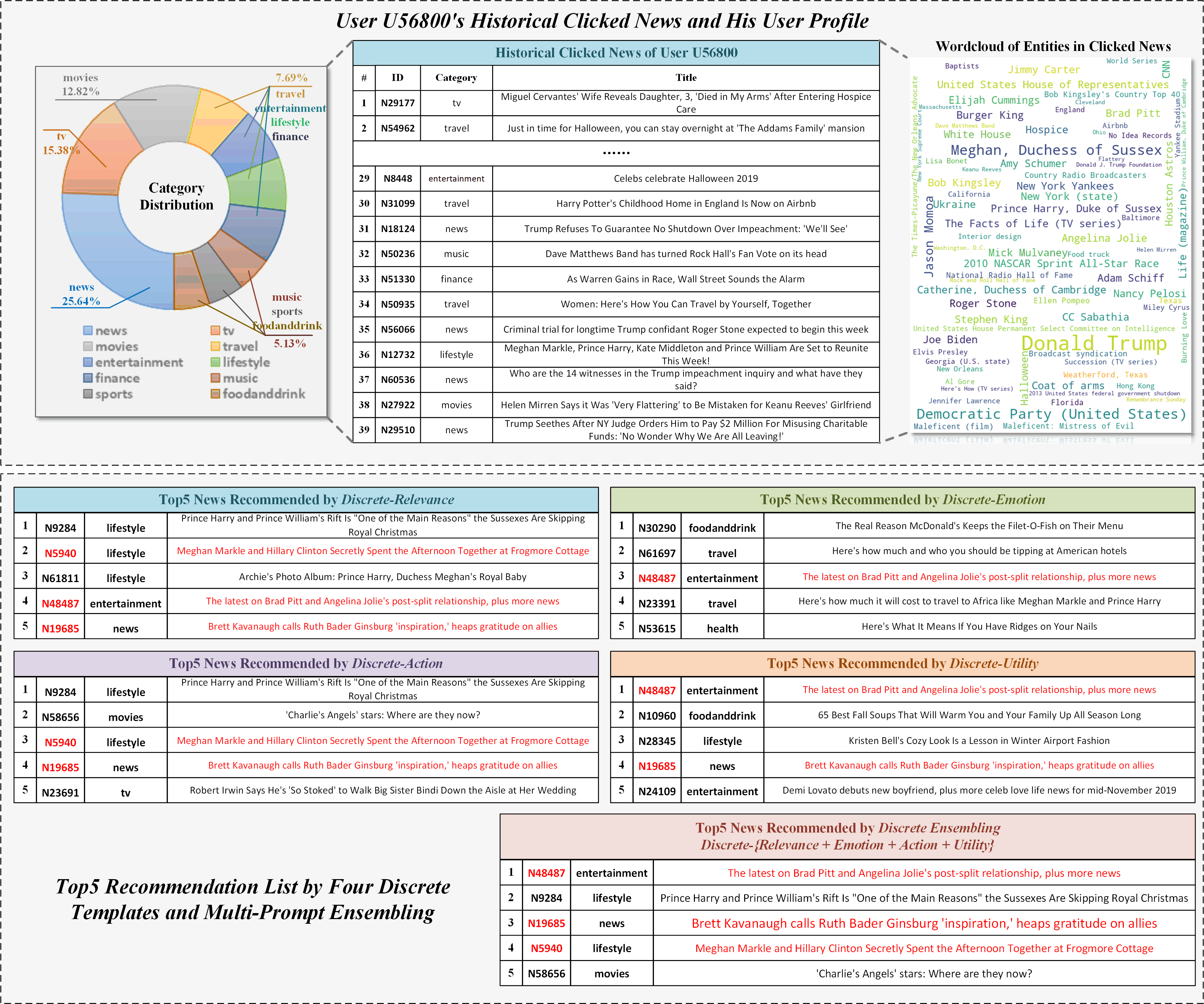}
	\caption{Case study of the top-5 news recommended by the four Discrete-$\{\text{Relevance, Emotion, Action, Utility}\}$ templates and Discrete-Ensembling for an impression of user \textit{U56800}. The recommended news actually clicked by the user is highlighted in red. The top is the historically clicked news and handmade profile of a user \textit{U56800}, including the category distribution and entity word-cloud of his history news. The bottom presents the five recommendation lists.}
	\label{Fig:CaseStudy-Rec}
\end{figure*}

\subsection{Case Study}
We conduct a case study to further intuitively shed light on the effectiveness of our Prompt4NR. Figure~\ref{Fig:CaseStudy-Rec} illustrates the top5 recommendation list formed by four one-single \textit{Discrete-$\{\text{Relevance, Emotion, Action, Utility}\}$} template and a \textit{Discrete-Ensembling} in an impression. In this impression, it consists of 39 historical news clicked by the user \textit{U56800} and 60 candidate news for ranking\footnote{These identifiers presented in the case study can be directly matched to the raw dataset, such as \textit{Uxxxx}, \textit{Nxxxx}.}. For more intuitive presentation, we mark the category label of news. Due to space consideration, not all 39 history news are shown in details, but we manually make a coarse user profile, including a category distribution (i.e., pie chart) and an entity frequency distribution (i.e., in the word-cloud, the higher frequency, the larger the word size) of his history news. From this case, we have following observations:

\par
This user has clicked news covering 10 categories. From the wordcloud of the user's clicked entities, words like "\textit{Meghan, Duchess of Sussex, Prince Harry, Duke of Sussex, Duchess of Cambridge}" are frequently in sight, and we can intuitively guess that the royal lifestyle may be one of the user's interests, but not the only one. In other words, this user's areas of interest are diverse. Besides, these candidate news are from multiple fields covering 14 categories (not plot due to space) and involve with numerous different entities. These observations imply that making an accurate recommendation for this user may not be an easy task.

\par
Although not easy, it is observed that our five recommendation lists all contain the ground-truth of user actually clicked news, simply called as \textit{true news}. We notice that each template has its own style of recommendation list. For example, the \textit{Discrete-Relevance} hits more true news, but less diversity. The \textit{Discrete-Emotion} recommends five news belonging to five different categories, even the "health" news never appears in the user's history. The \textit{Discrete-Action} and \textit{Discrete-Utility} both hit two true news: But the \textit{Discrete-Action} tends to the user's interest of royal lifestyle, and \textit{Discrete-Utility} finds another interest point of "entertainment" news. These results prove our motivation that though different templates seemingly with only a few differences, they may pay attention on different knowledge embedded in a PLM. Furthermore, it is observed that the \textit{Discrete-Ensembling} achieves a win-win situation compared with each one-single template. Specifically, the \textit{Discrete-Ensembling} also hits three true news, and their ranking positions are more beneficial: Five recommended news belonging to four categories presents diverse news to the user. This is an indication that the multi-prompt ensembling has the ability to incorporate the advantages of each one-single prompt for better recommendation.

\par
Incidentally, we notice an interesting point. The 20-th historical news \textit{N31748} is "Trailer - Charlie's Angels" about movies. In the top5 news recommended by the \textit{Discrete-Action}, there is a news \textit{N58656} "'Charlie's Angels' stars: Where are they now?", which is very related to \textit{N31748}. Although \textit{N58656} is not the true news, we believe that this is a nice recommendation, which can guide the user to review his potential interest and make a click action. This is a reflection of the Prompt4NR  capability to self-discover and serialize latent relations between news.

\par
In addition, it is observed that the four one-single templates are not perfect for thoroughly mining the user's interests. For example, no sports news is recommended here, but there is a sport news \textit{N40094} of "Baker Mayfield quick to condemn teammate Myles Garrett for brawl, attack on Mason Rudolph" actually clicked by the user. This suggests us that our four one-single templates are not enough to cover all recommendation considerations, which motivates us to design more excellent templates in the future work.

\section{Conclusion}
\label{Sec:Conclusion}
In this paper, we have proposed the Prompt4NR, a prompt learning framework for news recommendation, which transforms the NR task as a cloze-task for the \textsf{[MASK]} prediction task. As the first trial work, we have conducted extensive experiments with a set of various types of prompt templates, including discrete, continuous and hybrid templates. Moreover, we have adopted the multi-prompt ensembling to incorporate advantages from different prompts. Experiment results have validate the superiority of our Prompt4NR over the state-of-the-art competitors.

\section{Acknowledgements}
\noindent
This work is supported in part by National Natural Science Foundation of China (Grant No: 62172167). The computation is completed in the HPC Platform of Huazhong University of Science and Technology.

\newpage 

\normalem
\bibliographystyle{ACM-Reference-Format}
\bibliography{reference}
\end{sloppypar}
\end{document}